\documentclass[prb,preprint,showkeys,superscriptaddress]{revtex4}

\usepackage{graphicx}		% Include figure files
\usepackage{dcolumn}		% Align table columns on decimal point
\usepackage{bm}				% bold math
\usepackage{amsmath}

\usepackage{amssymb}
\usepackage{amsfonts}
\usepackage{color}
\usepackage{enumitem}
\usepackage{textcomp}

\DeclareMathOperator{\sinc}{sinc}

\newcommand*{\Scale}[2][4]{\scalebox{#1}{$#2$}}%

\begin{document}

\title{Lumped circuit model for inductive antenna spin-wave transducers}

\author{Frederic Vanderveken}
\email[E-mail: ]{frederic.vanderveken@imec.be}
\affiliation{Imec, 3001 Leuven, Belgium}
\affiliation{KU Leuven, Departement Materiaalkunde, SIEM, 3001 Leuven, Belgium}
\author{Vasyl Tyberkevych}
\affiliation{Department of Physics, Oakland University, Rochester, Michigan 48309, USA}
\author{Giacomo Talmelli}
\affiliation{Imec, 3001 Leuven, Belgium}
\affiliation{KU Leuven, Departement Materiaalkunde, SIEM, 3001 Leuven, Belgium}
\author{Bart Sor\'ee}
\affiliation{Imec, 3001 Leuven, Belgium}
\affiliation{KU Leuven, Departement Elektrotechniek, TELEMIC, 3001 Leuven, Belgium}
\affiliation{Universiteit Antwerpen, Departement Fysica, 2000 Antwerpen, Belgium}
\author{Florin Ciubotaru}
\affiliation{Imec, 3001 Leuven, Belgium}
\author{Christoph Adelmann}
\email[E-mail: ]{christoph.adelmann@imec.be}
\affiliation{Imec, 3001 Leuven, Belgium}

\date{\today}

\begin{abstract}
We derive a lumped circuit model for inductive antenna spin-wave transducers in the vicinity of a ferromagnetic medium. The model considers the antenna's Ohmic resistance, its inductance, as well as the additional inductance due to the excitation of ferromagnetic resonance or spin waves in the ferromagnetic medium. As an example, the additional inductance is discussed for a wire antenna on top of a ferromagnetic waveguide, a structure that is characteristic for many magnonic devices and experiments. The model is used to assess the scaling properties and the energy efficiency of inductive antennas. Issues related to scaling antenna transducers to the nanoscale and possible solutions are also addressed.
\end{abstract}


\maketitle
\section{Introduction}

Ferromagnetic resonance (FMR) and spin waves are precessional collective excitations of the magnetization in ferro-, ferri-, or antiferromagnetic media.\cite{Kittel1948,Herring1951,Stancil2009} Historically, FMR has found applications in microwave resonators, oscillators, and filters based on Yttrium iron garnet (YIG).\cite{Helszajn1985,Ishak1988} In recent years, much research has been devoted to miniaturizing spin-wave and FMR-based devices and structures in the growing field of (nano-)magnonics, with potential applications in spintronic computation\cite{Spintronics,Roadmap,Krug2010,Khitun2011,Nikonov2013,Demokritov2013,Chumak2015,Mahmoud2020,Fischer2017,Talmelli2020} and scaled microwave analog systems.\cite{YWL_2013,DBW_2020,Costa2021}

In magnonic devices, FMR or spin waves are excited by microwave electrical signals, employing transducers between electric (microwave) and magnetic (spin wave) domains. With alternatives based on spin-orbit torques\cite{Talmelli2018} or magnetoelectrics\cite{Weiler2011,Cherepov2014,Verba2018} only emerging, the vast majority of devices employs inductive antennas as transducers. Inductive antennas excite FMR or spin waves using oscillating Oersted magnetic fields generated by microwave currents and can also be utilized as a spin-wave detectors that inductively convert spin waves or FMR to microwave signals.

For practical applications, the power transmission from the peripheral microwave circuit into the magnetic system is of great importance. Historically, microstrip waveguides in combination with insulating ferrite films have been used typically in spin-wave devices. For such devices, distributed models of have been formulated that describe quantitatively the change in microwave impedance of the microstrip due to the excitation of FMR or spin waves.\cite{Stancil2009} Recently, micro- and nanomagnonic devices have however frequently used wire or U-shaped inductive antennas connected to coplanar waveguides since such structures can be more easily miniaturized.\cite{Fischer2017,Costa2021,Ciubotaru2016,Talmelli2021, Demidov14, Evelt16, Sebastian12, Sekiguchi10, Demidov09, Rousseau15, Sato13, Vogt14} For such antennas, the microwave behavior is better described by a lumped impedance of the inductive antenna and represented by an equivalent circuit. In this framework, the coupling to FMR and the spin-wave system can be described by a radiation impedance, in an analogous way to conventional electromagnetic antennas.\cite{Elliott2003,Balanis2016} The equivalent circuit can be used to quantitatively assess the spin-wave excitation and as a starting point for the design and optimization of inductive antennas. Thus, equivalent circuits are crucial, when such inductive antennas are to be included in spin-wave-based microwave or spintronic (logic) systems. This approach can ultimately be used in SPICE-like device models to simulate the behavior of magnonic or spintronic systems,\cite{Dutta2014, Connelly2021} including peripheral microwave circuits.

Several works have studied the radiation impedance of inductive antennas for particular cases. Ganguly \emph{et al.} first derived the spin-wave resistance and reactance for magnetostatic surface spin waves excited by a microstrip with uniform current density by solving the Maxwell equations with appropriate boundary conditions in all layers.\citep{Ganguly1975,Ganguly1978} Subsequent work has addressed gradually more complex structures and presented alternative derivations. Emtage reported a derivation based on surface permeabilities\cite{Emtage1978} whereas Kalinikos introduced a Green's function method to determine the spin-wave power and impedance.\cite{Kalinikos1981} Also non-uniform current densities were considered, \cite{Sethares1979,Emtage1982,Bajpai1988,Timoshenko2009, Dmitriev1988} as well as finite dimensions of the magnetic medium.\cite{Sethares1979,Kalinikos1981,Adam1982,Bajpai1988,Stancil2009} Furthermore, the spin-wave radiation impedance has also been derived for forward volume spin waves \cite{Emtage1978,Kalinikos1981,Weinberg1982,Adam1982,Yashiro1988,Dmitriev1988,Stancil2009,Vlaminck2010} as well as backward volume waves.\cite{Weinberg1982,Dmitriev1988,Timoshenko2009,Stancil2009,Vlaminck2010,Kostylev2016} Nevertheless, all these approaches require numerical methods to find the impedance and do not allow for simple approximate expressions that can be implemented in electrical models and equivalent circuits.

This paper presents an analytical compact lumped element model of an inductive antenna spin-wave transducer emitting dipolar-exchange spin waves. The model includes magnetic relaxation and can describe arbitrary static magnetization orientations. It allows for a complete representation of the transducer as an electric element, which can \emph{e.g.} be used in microwave circuit simulations to improve the impedance matching with a connected waveguide. The model is then used to study the consequences of scaling the dimensions of an inductive wire antenna and the influence of the magnetic material parameters. Finally, the maximum power transfer from the microwave to the magnetic domain is quantified for an inductive wire antenna, and implications of the wire antenna and waveguide scaling on the maximum power transfer are discussed.

\section{Lumped equivalent circuit model of inductive antenna spin-wave transducers}

The paper is organized as follows: this section presents the derivation of the equivalent impedance of an arbitrarily shaped antenna that is placed in the vicinity of a magnetic material. Thereafter, the impedance model is applied in Sec. III to the special cases of a magnetic thin film or waveguide and a straight wire antenna, configurations that are commonly used in magnonic experiments. Section~IV examines the impedance spectra and the scaling behavior of wire antennas on top of a CoFeB waveguide as a concrete example. Finally, Sec.~V discusses the optimum power transfer efficiency from the microwave to the magnetic domain, including for the above special case, and the dependence on antenna and waveguide dimensions.

\subsection{General geometry and model assumptions}

We consider a ferro- or ferrimagnetic medium, in which FMR or spin waves can be excited. The static magnetization of the medium and the static magnetic bias field are assumed to be parallel. Their orientation with respect to the geometry of the structure and the direction of spin-wave propagation can be variable and strongly determines the spin-wave properties. The model presented here is valid for all configurations. Several common specific cases are discussed below.

FMR or spin waves are excited by an inductive antenna with arbitrary shape and current distribution. The microwave current in the antenna generates an oscillating Oersted magnetic field, which interacts with the magnetic layer and excites the magnetization dynamics, \emph{i.e.} FMR or spin waves. Hence, there is a net energy flow from the microwave current inside the antenna to the magnetic medium. Our model treats spin waves as linear excitations of the magnetization and therefore neglects nonlinear effects. Spin waves in the linear regime are weak perturbations of the magnetization and thus carry small amounts of energy. In magnonic device applications, antenna transducers convert electrical signals to the magnetic domain (and \emph{vice versa}). For ultralow power applications, large energy flows into the spin-wave system are neither required nor desirable. However, even for signal conversion, high transducers efficiencies are essential to avoid large losses during transduction and the associated large external power requirements of the devices.  

For simplicity, the antenna length $\ell$ is assumed to be much smaller than the wavelength of an electromagnetic wave at the excitation frequency, \emph{i.e.} $\ell \ll \lambda_\mathrm{EM}$. In this limit, the phase of the electric current density along the antenna is constant and the transducer can be described by a lumped equivalent circuit rather than by a distributed-element model. Hence, the active and reactive parts of the energy flow can be captured in a lumped equivalent impedance.

\subsection{Impedance of an inductive antenna}

A microwave (radio-frequency, RF) current $I$ inside an inductive antenna generates an oscillating magnetic Oersted field $\bm{H}_\mathrm{a}(\bm{r})=\bm{h}_\mathrm{a}(\bm{r})I$ that can excite both resonant and nonresonant (evanescent) spin waves in an adjacent magnetic medium. For typical (nano-)magnonic devices, the Oersted field can be calculated in the magnetostatic approximation, which is valid when the structure's dimensions and the spin-wave wavelength are much smaller than the electromagnetic wavelength in vacuum ($\sim 10$ cm at GHz frequencies). Then, the influence of the time-varying electric field in the Maxwell equations can be neglected and the magnetic field can be calculated by the Biot--Savart law. For an arbitrary current distribution $\bm{J}(\bm{r}) = \bm{j}(\bm{r})I$, the magnetic field distribution $\bm{h}_\mathrm{a}$ can then be written as
\begin{align}
	\bm{h}_\mathrm{a}(\bm{r}) &= \frac{1}{4\pi} \int \bm{j}(\bm{r}')\times \frac{\bm{r}- \bm{r}'}{|\bm{r}-\bm{r}'|^3}\,d^3\bm{r}'\\
	\label{eq:Biot}
	&= \int \nabla_{\bm{r}} G(\bm{r},\bm{r}') \times  \bm{j}(\bm{r}')\,d^3\bm{r}'\, ,
\end{align}
\noindent with 
\begin{equation}
\label{eq:Green}
G(\bm{r},\bm{r}')=  \frac{1}{4\pi}\frac{1}{|\bm{r}-\bm{r}'|}
\end{equation}
the Green's function of the Laplace equation $\nabla^2_{\bm{r}} G(\bm{r},\bm{r}')=-\delta(\bm{r}-\bm{r}')$ and $\delta(\bm{r})$ the Dirac delta function. For simple geometries, the magnetic field distribution generated by the antenna $\bm{h}_\mathrm{a}$ can also be obtained directly from Amp\`ere's law, $\nabla_{\bm{r}}\times\bm{h}_\mathrm{a}(\bm{r})=\bm{j}(\bm{r})$ together with symmetry considerations, without the use of Eq.~\eqref{eq:Biot}. Below, in the description of the magnetization dynamics, we assume that $\bm{h}_\mathrm{a}(\bm{r})$ is known. The feedback of the magnetization dynamics (FMR or spin waves) on the antenna current distribution is thus neglected in this model. 

In addition to the Oersted magnetic field generated by the inductive antenna, magnetization dynamics (\emph{e.g.} FMR or spin waves) also generate a time-varying dipolar magnetic field. Both fields influence the voltage across the antenna via electromagnetic induction, \emph{i.e.} via an electromotive force (EMF). The impedance of the antenna can then be found by the ratio of the voltage across the antenna and the current, \emph{i.e.} $Z_\mathrm{eq}=V/I$. The total voltage drop is the sum of EMF contributions and the Ohmic resistance of the antenna. Hence, its equivalent impedance can be written as
\begin{equation}
\label{eq:Zeq}
	Z_\mathrm{eq} = R_\Omega + \frac{V_\mathrm{ind}}{I}\, ,
\end{equation}
with $R_\Omega$ the Ohmic antenna resistance and $V_\mathrm{ind}$ the EMF-induced voltage. Here, we neglect electromagnetic radiation losses in the nonmagnetic surroundings, which is valid in the magnetostatic limit. However, these can be added to the model as an additional impedance contribution.\cite{Elliott2003,Balanis2016}

The EMF-induced voltage originates from the time-varying magnetic fluxes generated by the Oersted and dipolar magnetic fields due to FMR or spin waves. By applying Stokes' theorem to the Maxwell--Faraday equation, the induced voltage at angular frequency $\omega = 2\pi f$ can be written as a function of the total flux, leading to 
\begin{equation}
	V_\mathrm{ind}= \oint_{C_a} \bm{E}(\bm{r})\, d\bm{l} = -i \omega \int_{S_a} \bm{B}(\bm{r})\, d^2\bm{a}  =-i\omega \Phi ,
\end{equation}
\noindent with $S_a$ the surface enclosed by the current loop, $\Phi$ the magnetic flux through this loop, $C_a = \partial S_a$, and $\bm{B}$ the magnetic induction. Considering the magnetic vector potential $\bm{A}$ defined by $\bm{B} = \nabla \times \bm{A}$, the magnetic flux can be written as 
\begin{equation}
	\Phi = \oint_{C_a} \bm{A}(\bm{r}) \, d\bm{l}\,.
\end{equation}
\noindent For a current distribution $\bm{J}(\bm{r}) = \bm{j}(\bm{r})I$, the line integral can be converted to a volume integral by averaging the magnetic vector potential weighted by the current density. This results in
\begin{align} \nonumber
\Phi &=  \int \bm{A}(\bm{r}) \cdot \bm{j}(\bm{r})\, d^3\bm{r}  \\ \nonumber
&= \iint \bm{A}(\bm{r}) \cdot \bm{j}(\bm{r}') \delta(\bm{r}-\bm{r}')\,  d^3\bm{r}\,  d^3\bm{r}'\\ \nonumber
&= -\iint \bm{A}(\bm{r}) \cdot \bm{j}(\bm{r}') \nabla^2_{\bm{r}} G(\bm{r},\bm{r}')\,  d^3\bm{r}\,  d^3\bm{r}' \\
\label{eq:flux}
&= -\iint   \nabla^2_{\bm{r}} \bm{A}(\bm{r}) \cdot \bm{j}(\bm{r}')  G(\bm{r},\bm{r}')\,  d^3\bm{r}\,  d^3\bm{r}'\,.
\end{align}
In the Lorenz gauge, the electrical current and the curl of the magnetization $\bm{M}$ can be interpreted as sources of the magnetic vector potential that satisfies the Poisson equation
\begin{equation}
	 \nabla^2_{\bm{r}} \bm{A}(\bm{r}) = -\mu_0 \left[\bm{j}(\bm{r})I +\nabla_{\bm{r}} \times \bm{M}(\bm{r})\right] \,.
\end{equation}
\noindent After substituting this into Eq.~\eqref{eq:flux}, the total flux $\Phi$ can be written as the sum of the flux originating from the current $\Phi_0$ and the flux originating from the curl of the magnetization $\Phi_m$
\begin{equation}
	\Phi = \Phi_0 + \Phi_m = L_0 I + L_mI \,,
\end{equation}
with $L_0$ the conventional self-inductance of the antenna and $L_m$ the additional inductance generated by FMR or spin waves. The self-inductance is given by
\begin{equation}
	L_0 = \mu_0\iint \bm{j}(\bm{r}) \cdot \bm{j}(\bm{r}')  G(\bm{r},\bm{r}')\,  d^3\bm{r}\,  d^3\bm{r}'
\end{equation}
\noindent and can be readily calculated when the current distribution is known. To obtain an expression for $L_m$, the magnetic flux needs to be considered. Integration by parts and substitution of Eq.~\eqref{eq:Biot} leads to
\begin{align} \nonumber
	\Phi_m &= \mu_0 \iint \left[\nabla_{\bm{r}} \times \bm{M}(\bm{r}) \right] \cdot \bm{j}(\bm{r}')  G(\bm{r},\bm{r}')\,  d^3\bm{r}\,  d^3\bm{r}' \\ \nonumber
	&= \mu_0 \iint  \bm{M}(\bm{r})  \cdot \left[\nabla_{\bm{r}} \times \bm{j}(\bm{r}')  G(\bm{r},\bm{r}')\right]\,  d^3\bm{r}\,  d^3\bm{r}' \\ \nonumber
	&= \mu_0 \int  \bm{M}(\bm{r})  \cdot \left[\int\nabla_{\bm{r}} G(\bm{r},\bm{r}')\times \bm{j}(\bm{r}')   d^3\bm{r}'\right]\,  d^3\bm{r}  \\
	\label{eq:Fm}
	&= \mu_0 \int  \bm{M}(\bm{r})  \cdot \bm{h}_\mathrm{a}(\bm{r})\,  d^3\bm{r}  \,.
\end{align}

For linear magnetization dynamics described by the linearized Landau–-Lifshitz–-Gilbert equation, the dynamic magnetization can be written as \cite{Kalinikos1986}
\begin{align} \nonumber
	 \bm{M}(\bm{r})&=\int \hat{\chi}_\omega(\bm{r},\bm{r}') \bm{H}_\mathrm{a}(\bm{r}')\, d^3\bm{r}' \\
	 \label{eq:sus}
	 &= I\int \hat{\chi}_\omega(\bm{r},\bm{r}')\bm{h}_\mathrm{a}(\bm{r}')\, d^3\bm{r}'\,.
\end{align}
\noindent Here, $\hat{\chi}_\omega(\bm{r},\bm{r}')$ is the dynamic magnetic susceptibility tensor and $\bm{h}_\mathrm{a}$ the dynamic Oersted field. Substitution of Eq.~\eqref{eq:sus} into Eq.~\eqref{eq:Fm} results in an excess inductance generated by FMR or spin waves that is given by
\begin{equation}
	L_m =\frac{\Phi_m}{I} = \mu_0 \iint \bm{h}^T_\mathrm{a}(\bm{r}')\hat{\chi}_\omega(\bm{r},\bm{r}')  \bm{h}_\mathrm{a}(\bm{r})\, d^3\bm{r}\, d^3\bm{r}'\, .
\end{equation}
\noindent This is the general expression to determine the spin-wave generated partial inductance introduced in an arbitrary shaped conductive wire. Hence, to find $L_m$, one needs to know the antenna's field distribution and the spin-wave susceptibility. Note that the self-inductance can also be expressed in a similar form
\begin{equation}
L_0 = \mu_0 \iint \bm{h}^T_\mathrm{a}(\bm{r}')\hat{\chi}_0(\bm{r},\bm{r}')  \bm{h}_\mathrm{a}(\bm{r}) d^3\bm{r}d^3\bm{r}'
\end{equation}
with $\hat{\chi}_0(\bm{r}-\bm{r}')=\delta(\bm{r}-\bm{r}') \bm{I}$ the vacuum susceptibility and $\bm{I}$ the identity matrix.

\begin{figure}[tb]
	\includegraphics[width=8.5cm]{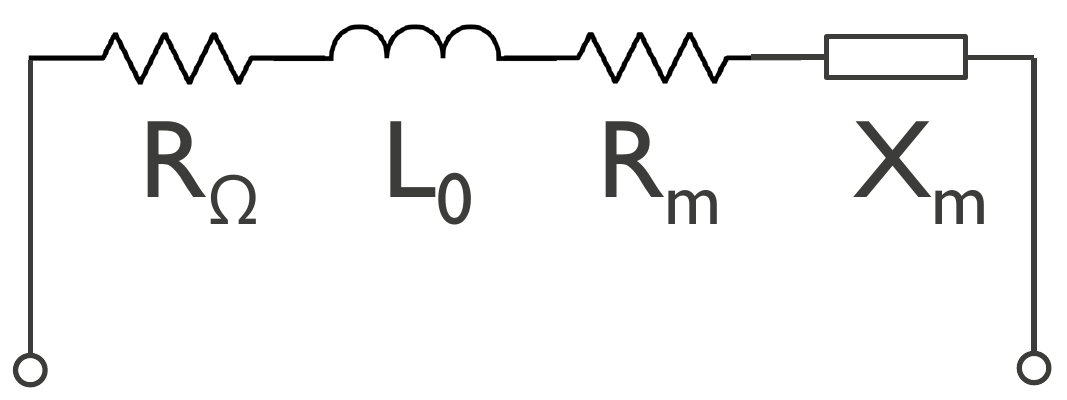}
	\caption{Lumped equivalent circuit representation of an inductive antenna spin-wave transducer. The equivalent circuit consists of an Ohmic series resistance ($R_\Omega$), a self-inductance ($L_0$), as well as a radiation resistance ($R_m$) and reactance ($X_m$) that capture the power coupling to FMR or spin waves.}
	\label{fig:eq}
\end{figure}

Based on this model, an inductive antenna can therefore be represented by a circuit comprising its Ohmic resistance $R_\Omega$, its self-inductance $L_0$, as well as an additional partial inductance $L_m$ that stems from the coupling to the magnetic system. As shown below, $L_\mathrm{m} = L_\mathrm{m}'+iL_\mathrm{m}''$ is in general complex because the dynamic magnetic susceptibility tensor $\hat{\chi}_\omega(\bm{r},\bm{r}')$ is complex. The impedance $Z_m = i\omega L_m$ can therefore be divided into a resistance $R_m = -\omega L_\mathrm{m}''$ and a reactance $X_m = \omega L_\mathrm{m}'$. A schematic of the equivalent circuit is shown in Fig.~\ref{fig:eq}. In this equivalent circuit, the radiation resistance $R_m$ describes the dissipation due to spin-wave or FMR excitation and can therefore be used to describe the energy transfer between electric and magnetic domains. This will be used below to discuss the energy efficiency and the scaling behavior of inductive antennas.

\section{Inductive antenna on top of a ferromagnetic waveguide with uniform magnetization dynamics}

\subsection{Additional inductance due to spin-wave or FMR excitation}

An inductive antenna on top of a ferro- or ferrimagnetic thin-film waveguide is a common basic structure in many (nano-)magnonic experiments. We therefore apply the above general model to the case of a magnetic waveguide with thickness $t$. The spin waves are assumed to propagate along the $x$-axis, whereas the $z$-axis is chosen perpendicular to the waveguide surface. We assume that the waveguide thickness $t$ is much smaller than the spin-wave wavelength $\lambda$, \emph{i.e.} $kt\ll1$, with $k = 2\pi /\lambda$, as this results in uniform magnetization dynamics over the thickness. Furthermore, we also assume that there is no lateral mode formation, and the magnetization dynamics are uniform over the width of the waveguide. This is realized in narrow waveguides when the spin-wave wavelength is much larger than the waveguide width. For wider waveguides, the boundary conditions at its edges leads to the formation of spin-wave modes with nonuniform magnetization dynamics over the waveguide width, which are difficult to treat analytically. However, for first-order spin-wave modes in sufficiently wide waveguides (with extended thin films as a limit), edge effects become again negligible, and the magnetization dynamics become approximately uniform. Hence, the analytical calculations are appropriate both for narrow and wide waveguides with widths $W$ so that $kW \ll 1$ or $kW \gg 1$. For waveguides with $kW \sim 1$, mode formation cannot be neglected, and numerical calculations are required to assess the inductance of an inductive antenna.

For uniform magnetization dynamics, the integration over the width and thickness directions can be replaced by a multiplication with the structure's cross section
\begin{equation}
\label{eq:Lm_general}
L_m = \mu_0 tW \iint \bm{h}^T_\mathrm{a}(x')\hat{\chi}_\omega(x,x')  \bm{h}_\mathrm{a}(x)\, dx\, dx' \, ,
\end{equation}
\noindent with $t$ the thickness and $W$ the width of the waveguide.

The result can be simplified by writing the expression in reciprocal space. The dynamic susceptibility is translationally invariant, \emph{i.e.} it only depends on the distance between two points $\bm{r}$ and $\bm{r}'$, and can therefore be written as $\hat{\chi}_\omega(\bm{r},\bm{r}') = \hat{\chi}_\omega(\bm{r}-\bm{r}',0)$. Taking the Fourier transform, applying Plancherel's theorem, and considering translational invariance results in 
\begin{equation}
\label{eq:Lm}
	L_m = \mu_0 t W \int \bm{h}^T_\mathrm{a}(k)\hat{\chi}_\omega(k)  \bm{h}^*_\mathrm{a}(k)\, \frac{dk}{2\pi} \, .
\end{equation}
\noindent For linear magnetization dynamics, the susceptibility in reciprocal space can be written as
\begin{equation}
\label{eq:chi}
	\hat{\chi}_\omega(k) = \frac{\hat{\Omega}(k)^2}{\omega_\mathrm{r}^2(k)-\omega^2 + i \omega \Gamma(k)}\,,
\end{equation}
\noindent with $\omega_\mathrm{r}(k) = 2\pi f_\mathrm{r}(k) $ the spin-wave dispersion relation, $\Gamma(k)$ the spin-wave damping rate, and $\hat{\Omega}(k)$ a tensor dependent on the magnetic bias field and wavenumber $k$. Analytical expressions of these quantities can be found in Appendix~\ref{app:A}. 

For low damping, \emph{i.e.} for Gilbert damping constants $\alpha \ll 1$, Eq.~\eqref{eq:chi} can be further approximated by 
\begin{equation}
\label{eq:chi_app}
\hat{\chi}_\omega(k) \approx \frac{\hat{\Omega}(k)}{\omega_\mathrm{r}(k)-\omega + i \Gamma(k)}\,.
\end{equation}
\noindent Using this approximation, an analytical solution to the integral in Eq.~\eqref{eq:Lm} can be found
\begin{equation}
\label{eq:Lm_final}
L_m = i\frac{\mu_0 t W}{|v_g(k_\omega)|}  \left[\bm{h}^T_\mathrm{a}(k_\omega)\hat{\Omega}_\omega(k_\omega)  \bm{h}^*_\mathrm{a}(k_\omega)\right]\, ,
\end{equation}
\noindent with $k_\omega$ the wavevector at the spin-wave resonance for angular frequency $\omega = 2\pi f$, \emph{i.e.} $\omega_\mathrm{r}(k_\omega) = \omega$, and $v_g(k)=\partial\omega_\mathrm{r}(k)/\partial k$ the spin-wave group velocity. Note that reciprocal spin-wave propagation was assumed in the integration, which means that both propagation directions of excited spin waves $\pm k_\omega$ provide equal contributions to $L_m$. This is a valid assumption if the waveguide thickness is much smaller than the spin-wave wavelength, \emph{i.e.} $k_\omega t\ll1$.

Equation \eqref{eq:Lm_final} also takes into account nonresonantly excited (evanescent) spin waves. Furthermore, unlike the self-inductance $L_0$, the spin-wave inductance $L_m$ is generally a complex quantity with both real and imaginary parts, \emph{i.e.} $L_\mathrm{m} = L_\mathrm{m}'+iL_\mathrm{m}''$. The real part leads to a reactance in the equivalent circuit, $X_\mathrm{m} = \omega L_\mathrm{m}'$, whereas the imaginary part corresponds to a resistance, $R_\mathrm{m} = -\omega L_\mathrm{m}''$, as discussed above. Combining the previous results in Eq.~\eqref{eq:Zeq} in the framework of the equivalent circuit in Fig. \ref{fig:eq} leads to an impedance of an antenna above a magnetic thin film or waveguide of
\begin{equation}
Z_\mathrm{eq} = R_\Omega - i\omega \left(L_0 +  L_m\right) = R_\Omega -  R_\mathrm{m} - i\omega (L_0 + L_\mathrm{m}') \, ,
\label{eq:z_antenna}
\end{equation}
with $L_m$ given by Eq.~\eqref{eq:Lm_final}.

\subsection{Two-antenna configurations: Mutual inductance due to propagating spin-waves\label{mutual}}

The above derived model can be extended to describe the mutual inductance between two inductive antennas connected via a magnetic waveguide. As explained above, the magnetization dynamics generates a dipolar magnetic field which induces an additional voltage in the antenna. The magnetization dynamics is not localized near the exciting inductive antenna but spreads over the magnetic medium. This can be interpreted as spin waves that propagate in the magnetic medium away from the source. A second ``receiving'' inductive antenna that is placed at some distance $D$ from the excitation antenna then also feels the dynamic dipolar field associated with the magnetization dynamics. Consequently, this field also induces a voltage in the receiving antenna. The interaction between the two antennas can be represented by a mutual inductance $\mathcal{M}$. The derivation of $\mathcal{M}$ is analogous to the above derivation of the additional magnetic inductance $L_m$. However, here, a space separation $D$ between the two antennas should be considered in equation Eq.~\eqref{eq:Lm_general}
\begin{equation}
\mathcal{M} = \mu_0 tW \iint \bm{h}^T_\mathrm{a}(x')\hat{\chi}_\omega(x,x')  \bm{h}_\mathrm{a}(x-D)\, dx\, dx' \, ,
\end{equation}
In reciprocal space, this leads to a phase accumulation of $kD$ during propagation and gives
\begin{align}
\label{eq:Lmutual}\nonumber
\cal{M} &= \mu_0 t W \int e^{ikD} \bm{h}^T_\mathrm{a}(k)\hat{\chi}_\omega(k)  \bm{h}^*_\mathrm{a}(k)\, \frac{dk}{2\pi}\\
&= \mu_0 t W e^{-k_iD} \int e^{ik_rD} \bm{h}^T_\mathrm{a}(k)\hat{\chi}_\omega(k)  \bm{h}^*_\mathrm{a}(k)\, \frac{dk}{2\pi}\, .
\end{align}
Here, $k = k_r + ik_i$ is the wavevector of the propagating spin wave. Effects of magnetic damping are captured by the imaginary component of the complex wavevector, resulting in an exponential decay of $\cal{M}$ with increasing inter-antenna distance $D$. We note that near FMR ($k \approx 0$), the mutual inductance becomes equal to the additional spin-wave induced inductance of the excitation antenna, i.e. ${\cal{M}} \approx  L_m$.

\subsection{Radiation impedance of a wire antenna on top of a ferromagnetic narrow waveguide with uniform magnetization dynamics}

\begin{figure}[tb]
	\includegraphics[width=8.5cm]{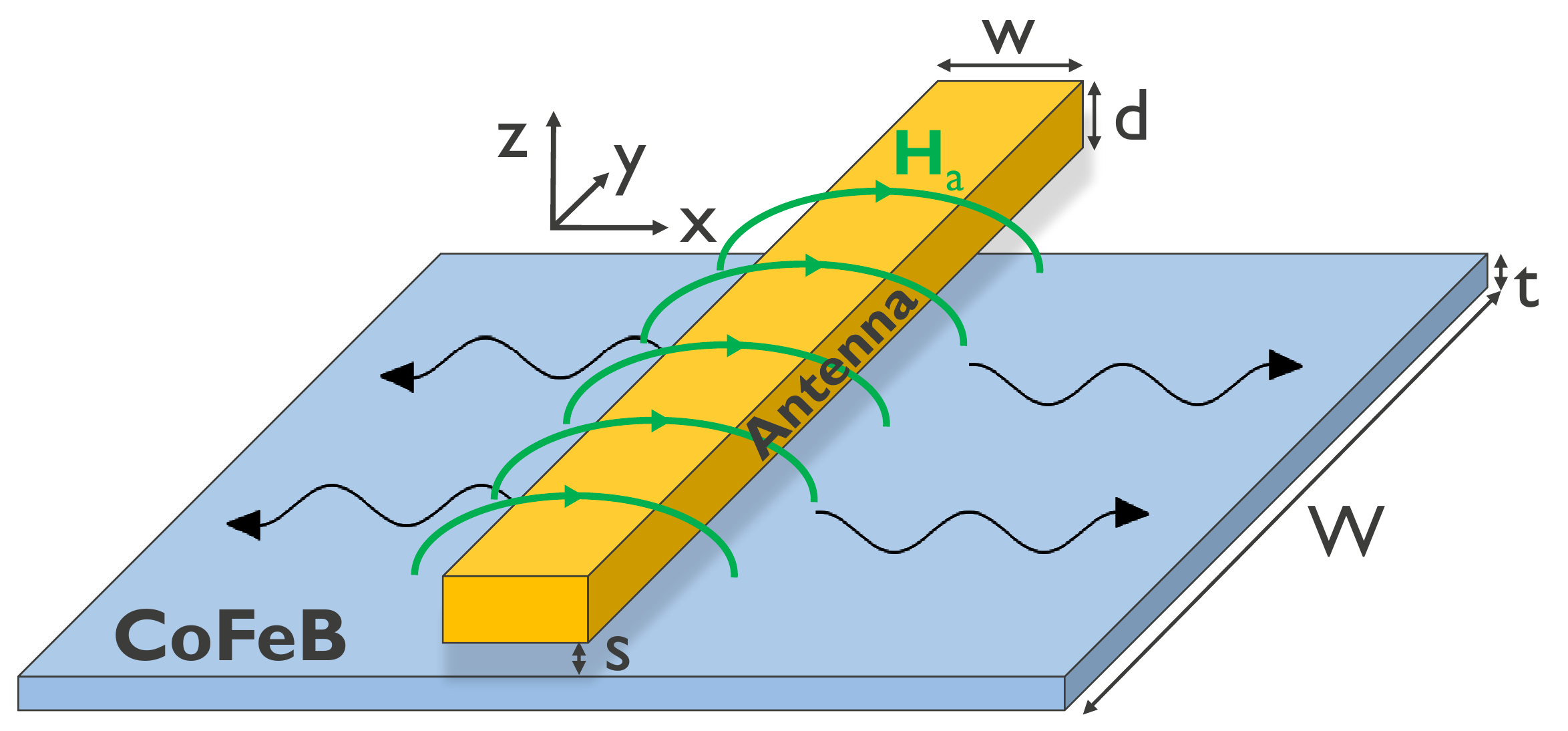}
	\caption{Schematic of the studied geometry: wire antenna (yellow) on top of a ferromagnetic narrow waveguide or thin film (blue). The static magnetization inside the waveguide or thin film can take any arbitrary direction in the general model.}
	\label{fig:Geometry}
\end{figure}

As a more concrete structure, we now consider a straight wire antenna with rectangular cross section at a distance $s$ above a magnetic thin film, as used in various spin-wave propagation experiments.\cite{Fischer2017,Costa2021,Ciubotaru2016,Talmelli2021, Demidov14, Evelt16, Sebastian12, Sekiguchi10, Demidov09, Rousseau15, Sato13, Vogt14} The structure and the geometry of the studied structure are depicted in Fig.~\ref{fig:Geometry}. To simplify the system, we assume a uniform current density inside the antenna, \emph{i.e.} we neglect the impact of the skin effect. This is justified as long as the antenna width or thickness are much smaller than the skin depth $\delta = \sqrt{\frac{\rho}{\pi f \mu}}$ at frequency $f$. Here, $\rho$ is the resistivity of the antenna metal and $\mu = \mu_0\mu_r$ its permeability. For Cu, the skin depth is of the order of 1 $\mu$m at GHz frequencies (larger for poorer conductors) and thus the approximation is acceptable for the nanoscale antennas considered below. The Ohmic resistance of the antenna is then described by Pouillet’s law
\begin{equation}
\label{eq:Rs}
	R_\Omega = \frac{\rho \ell}{dw},
\end{equation}
\noindent with $\ell$, $d$, and $w$ the antenna length, thickness, and width, respectively. If the skin effect cannot be neglected, an effective area $A_\mathrm{eff}$ based on the current distribution inside the antenna needs to replace $dw$ in Eq.~\eqref{eq:Rs}. The self-inductance for this geometry can be approximated by \cite{Clayton2010}
\begin{equation}
L_0 = \frac{\mu_0\ell}{2\pi} \left[ \ln\left( \frac{2\ell}{w+d}\right) +\frac{1}{2} \right]   \,.
\label{eq:L0}
\end{equation}

To evaluate the additional inductance $L_m$ in Eq.~\eqref{eq:Lm} that originates from the excitation of FMR or spin waves, the Oersted field of the antenna, $\bm{H}_\mathrm{a}$, needs to be determined. A simple analytical expression for $\bm{H}_\mathrm{a}$ can be found when two approximations hold: (i) the film thickness $t$ is smaller than the sum of the antenna thickness and the separation $d+s$ which, (ii) is in turn assumed to be smaller than the antenna width $w$, \emph{i.e.} $t\ll d+s \ll w$. For this geometry, the Oersted field inside the magnetic waveguide can be approximated by
\begin{equation}
\label{eq:Hoe}
\bm{H}_\mathrm{a}(\bm{r})  = \frac{I}{2(w+\pi s)}\Pi\left(\frac{x}{w}\right) \bm{\hat{e}}_{x}\, ,
\end{equation}
with $\Pi\left(x\right)$ the rectangular function (defined as 1 for $|x|\leq \frac{1}{2}$ and 0 for $|x|>\frac{1}{2}$) and $\bm{\hat{e}}_{x}$ the unit vector in the $x$-direction. The spatial Fourier transform of this field is given by
\begin{equation}
\label{eq:Hoe_k}
\bm{H}_\mathrm{a}(\bm{k}) = \frac{I}{k(w+\pi s)} \sin\left( \frac{k w}{2}\right)\bm{\hat{e}}_{k}\,.
\end{equation}
\noindent When $w\ll s$, the Oersted field distribution becomes a sinc-function
\begin{equation}
	\bm{h}_\mathrm{a}(\bm{k}) = \frac{1}{2} \sinc\left( \frac{k w}{2}\right)\bm{\hat{e}}_{k}\,.
\end{equation}
Using Eq.~\eqref{eq:Lm} and further assuming $s\ll w$, the partial spin-wave inductance can then be written as
\begin{equation}
\label{eq:Lmww}
	L_m = \mu_0 t W \int \chi_{\omega,xx}(k)  \left[\frac{1}{2}\sinc\left(\frac{k w}{2}\right)\right]^2\,\frac{dk}{2\pi}\,.
\end{equation}
\noindent Multiplying the inductance with the angular frequency results in the spin-wave impedance $Z_m = R_m + iX_m$ with the spin-wave resistance 
\begin{equation}
\label{eq:Rm}
	R_m = \frac{\omega \mu_0 t W}{8\pi} \int \chi_{\omega,xx}''(k)  \sinc^2\left(\frac{k w}{2}\right)\, dk
\end{equation}
and the spin-wave reactance
\begin{equation}
\label{eq:Xm}
	X_m = \frac{\omega \mu_0 t W}{8\pi} \int \chi_{\omega,xx}'(k)  \sinc^2\left(\frac{k w}{2}\right)\, dk\, ,
\end{equation}
\noindent with the complex susceptibility $\hat{\chi}_{\omega,xx}(k)=\hat{\chi}_{\omega,xx}'(k)- i\hat{\chi}_{\omega,xx}''(k)$ derived in Appendix~\ref{app:A}.

\section{Scaling behavior of the spin-wave impedance of inductive wire antenna transducers}

\subsection{Impedance spectra of inductive wire antenna transducers}

\begin{figure}[tb]
	\includegraphics[width=16cm]{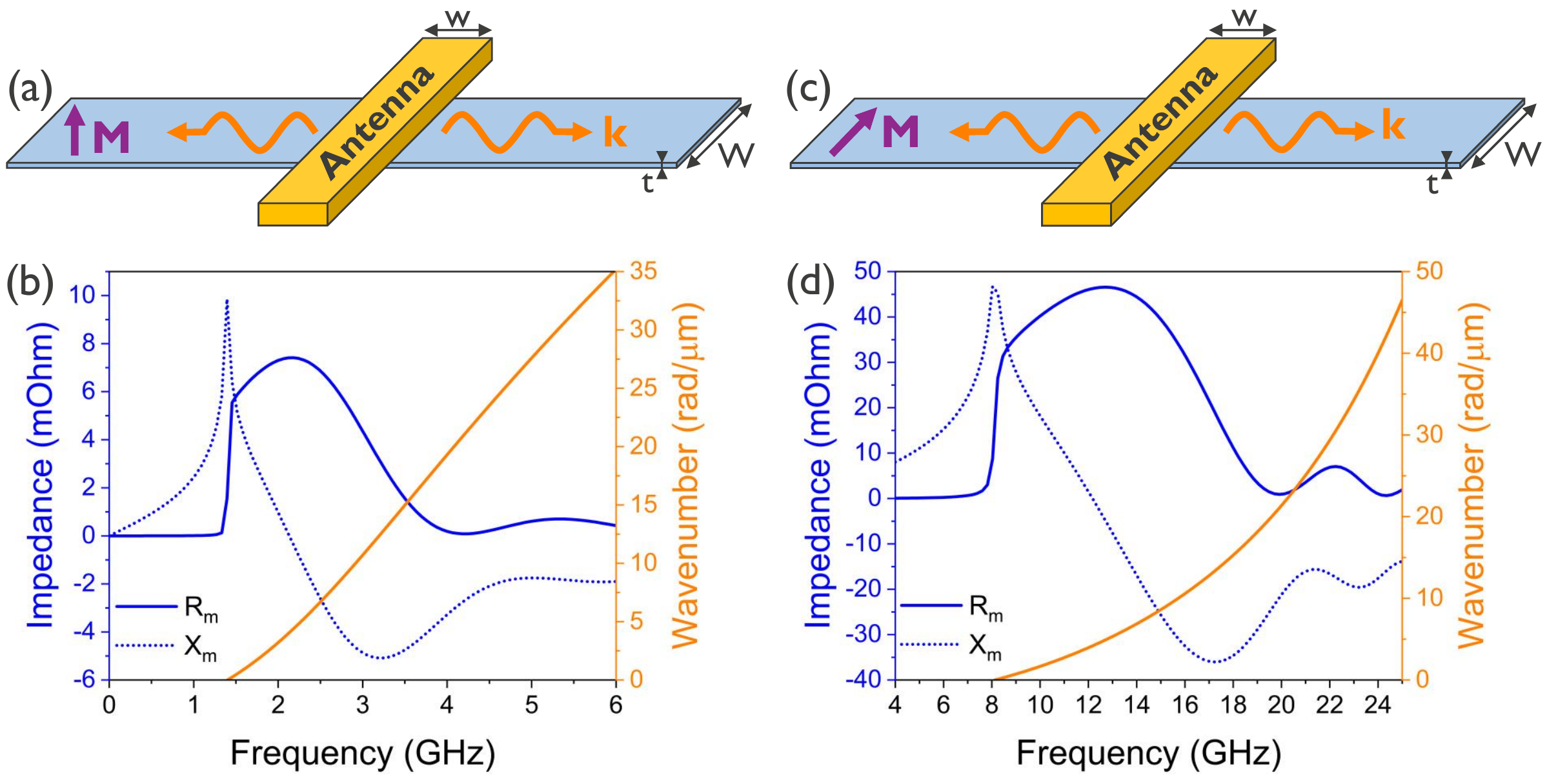}
	\caption{Spin-wave impedance of a 300 nm wide antenna on top of a 1 \textmu{}m wide and 20 nm thick CoFeB waveguide (see the text for material parameters) in the \textbf{(a)} Damon--Eshbach and \textbf{(c)} the forward-volume configurations. \textbf{(b)} and \textbf{(d)} show the real ($R_m$) and imaginary ($X_m$) components of the complex impedances as a function of frequency $f$ in the Damon--Eshbach and forward-volume configurations, respectively. In addition, the spin-wave dispersion relations are shown as yellow lines. The FMR frequency is located at zero wavenumber; for the parameters considered, it is 1.4~GHz and 8.1~GHz for the forward-volume and Damon--Eshbach configurations, respectively.}
	\label{fig:impedance}
\end{figure}

The spin-wave resistance $R_m$ and reactance $X_m$ in Eqs.~\eqref{eq:Rm} and \eqref{eq:Xm} are critical parameters that determine the measured signals in magnonic experiments using inductive wire antennas and ferri- or ferromagnetic waveguides. Their scaling behavior is key when magnonic devices are miniaturized to the micro- and nanoscale. We therefore discuss here the general behavior of $R_m$ and $X_m$ for a concrete example based on a thin CoFeB waveguide, assuming laterally uniform magnetization dynamics. Since both the spin-wave dispersion relations and the susceptibility $\hat{\chi}_{\omega,xx}(k)$ depend on the relative orientations between the static magnetization, the normal to the waveguide plane, and the spin-wave wavevector $\bm{k}$ that points along the waveguide, three different configurations need to be distinguished (\emph{cf.}~Fig.~\ref{fig:impedance}): (i) the backward-volume configuration with the magnetization and the wavevector both in-plane and parallel; (ii) the forward-volume configuration with the magnetization out-of-plane and the wavevector in-plane; and (iii) the Damon--Eshbach configuration with the magnetization and the wavevector both in-plane and perpendicular.

In the following, we discuss, as an example, the spectral characteristics of the spin-wave resistance $R_m$ and reactance $X_m$ of a wire antenna above a ferromagnetic CoFeB waveguide\cite{An14, Demidov14, Yua12, Wei19, Talmelli2021} with laterally uniform magnetization dynamics. The used material and geometric parameters are: saturation magnetization $M_s$ = 1.3 MA/m; exchange constant $A_\mathrm{ex}$ = 18 pJ/m; Gilbert damping $\alpha$ = 0.004; static bias field $\mu_0 H_0$ = 50 mT; waveguide width $W$ = 1 \textmu{}m; CoFeB thickness $t$ = 20 nm; and antenna width $w$ = 300 nm. We focus on the Damon--Eshbach and forward-volume configurations as spin waves are only efficiently excited in such geometries by an inductive antenna transducer. By contrast, the excitation efficiency of backward-volume spin waves\cite{Sato2014,Wessels2016,Bhaskar20} is much lower since the in-plane component of the antenna excitation field is parallel with the static magnetization and thus does not generate any torque. Only the much smaller out-of-plane component of the Oersted field can excite spin waves in the backward-volume configuration.\cite{Bhaskar20} While the above approach is also valid for the case where the $z$-component of the Oersted field is not neglected, we omit for simplicity the treatment of the backward-volume configuration in the following since the excitation efficiency for this geometry is zero in our approximation. 

The resulting spin-wave impedances in the two considered geometric configurations are plotted in Fig.~\ref{fig:impedance}. Several aspects can be identified from these plots, which are common to both configurations. Below the FMR frequency, the spin-wave resistance is zero, whereas the reactance remains nonzero for considerably smaller frequencies. This means that, below FMR, no net power transfer takes place from the electric to the magnetic domain. Instead, resonant power oscillations between the two domains occur that originate from evanescent spin waves. Near FMR (which is broadened by damping), the magnetic resonance conditions are met and net power transfer from the electric into the magnetic system occurs. At the same time, the reactance reaches a maximum at the FMR frequency. 

Above the FMR frequency, the resistance $R_m$ increases further until it reaches a maximum. The maximum is due to the interplay of three factors in Eq.~\eqref{eq:Rm}: the angular frequency $\omega$, the complex part of the susceptibility, and the spectrum of the Oersted field generated by the antenna. At frequencies just above FMR, the susceptibility and Oersted field spectrum are quasi-constant and the resistance increases linearly with frequency, as it is proportional to $\omega$ (\emph{cf.}~Eq.~\eqref{eq:Lm}). At higher frequencies, and thus also at higher wavenumbers, the Oersted field spectrum dominates. This results in strongly damped oscillations and an overall decrease of the resistance at higher frequencies.

By contrast, the reactance $X_m$ is strongly reduced above the FMR frequency until it reaches a negative minimum. This decrease with frequency can be attributed to the real part of the susceptibility. At a specific frequency, \emph{i.e.} the frequency for which $\chi'(k) = 0$, the reactance becomes zero and no evanescent spin waves are generated. Below this frequency, the reactance is positive and has inductive behavior, whereas above this frequency the reactance is negative and has capacitive behavior. At even higher frequencies, the Oersted field spectrum also becomes dominant for the reactance, which results in damped oscillations and an overall decay.

A comparison of Figs.~\ref{fig:impedance}(a) and \ref{fig:impedance}(b) reveals that the impedances for both forward-volume and Damon--Eshbach spin waves share the general trends. The main difference lies in the resonance frequency (FMR), which has a value of 1.4~GHz in the forward-volume and 8.1~GHz in the Damon--Eshbach configuration, as well as in the magnitude of the spin-wave impedance. For identical CoFeB parameters, the spin-wave resistance and reactance are about an order of magnitude larger for the Damon--Eshbach configuration than for the forward-volume configuration which is due to differences in the respective susceptibilities and the higher working frequencies in the Damon--Eshbach configuration. We remark that impedance values in the backward-volume configuration are again more than one order of magnitude lower than in forward-volume configuration due to the weak $z$-component of the Oersted field that is required to excite spin waves. 

\subsection{Scaling behavior}

In recent years, numerous experiments have been conducted to study spin waves in nanoscale magnetic structures.\citep{Talmelli2021, Vanderveken20, Talmelli2020, Mohseni20, Wang20} When such spin waves are excited by inductive antennas, the scaling behavior of the spin-wave impedance of the antenna--waveguide system is key to understand the experimental signals and their dependence on the device geometry and dimensions. The scaling behavior of the system can be divided in two parts: (i) the dependence of the spin-wave impedance on the antenna and waveguide dimensions and (ii) the power transfer between the electrical source and the spin-wave system, which depends on the entire equivalent circuit, as represented in Fig.~\ref{fig:eq}. Here, we first discuss (i) whereas (ii) will be addressed in the next section. Since the spin-wave impedance is much larger in the Damon--Eshbach configuration than in the forward-volume configuration, we focus in the following on the former. However, the above results indicate that the two configurations share general trends, the conclusions are also qualitatively valid for spin waves in the forward-volume geometry. All results are again based on the CoFeB material parameters listed above.

The three main geometrical parameters of the antenna--waveguide system that influence the spin-wave impedance are the magnetic waveguide thickness $t$, the waveguide width $W$, and the antenna width $w$. Equation~\eqref{eq:Lm} indicates that, for plane waves, the impedance is simply proportional to the waveguide width $W$ as is the case for very wide waveguides and narrow waveguides with a waveguide width in the order of the exchange length. For intermediate cases with $W$ comparable to the spin-wave wavelength $\lambda$, the lateral confinement of the spin waves leads to mode formation, which complicates the expressions for the susceptibility and spin-wave excitation efficiency. In this case, the spin-wave impedance and the influence of the waveguide width typically needs to be determined numerically. Nevertheless, some qualitative predictions can be made by considering the overlap integral approach to determine the excitation efficiency of a particular mode. The higher the overlap between the Oersted excitation field and the spin-wave mode, the stronger the excitation and thus the larger the spin-wave resistance. Hence, increasing the waveguide width results in a higher volumetric overlap integral and thus higher spin-wave resistances. The influence of the mode profile can be qualitatively captured by a form factor that takes into account the effective mode amplitude, for example the profile rms-value instead of its peak value. Therefore, the mode formation typically results in a slightly smaller net overlap integral as compared to the overlap integral with a plane wave. 

\begin{figure*}[tb]
	\includegraphics[width=16cm]{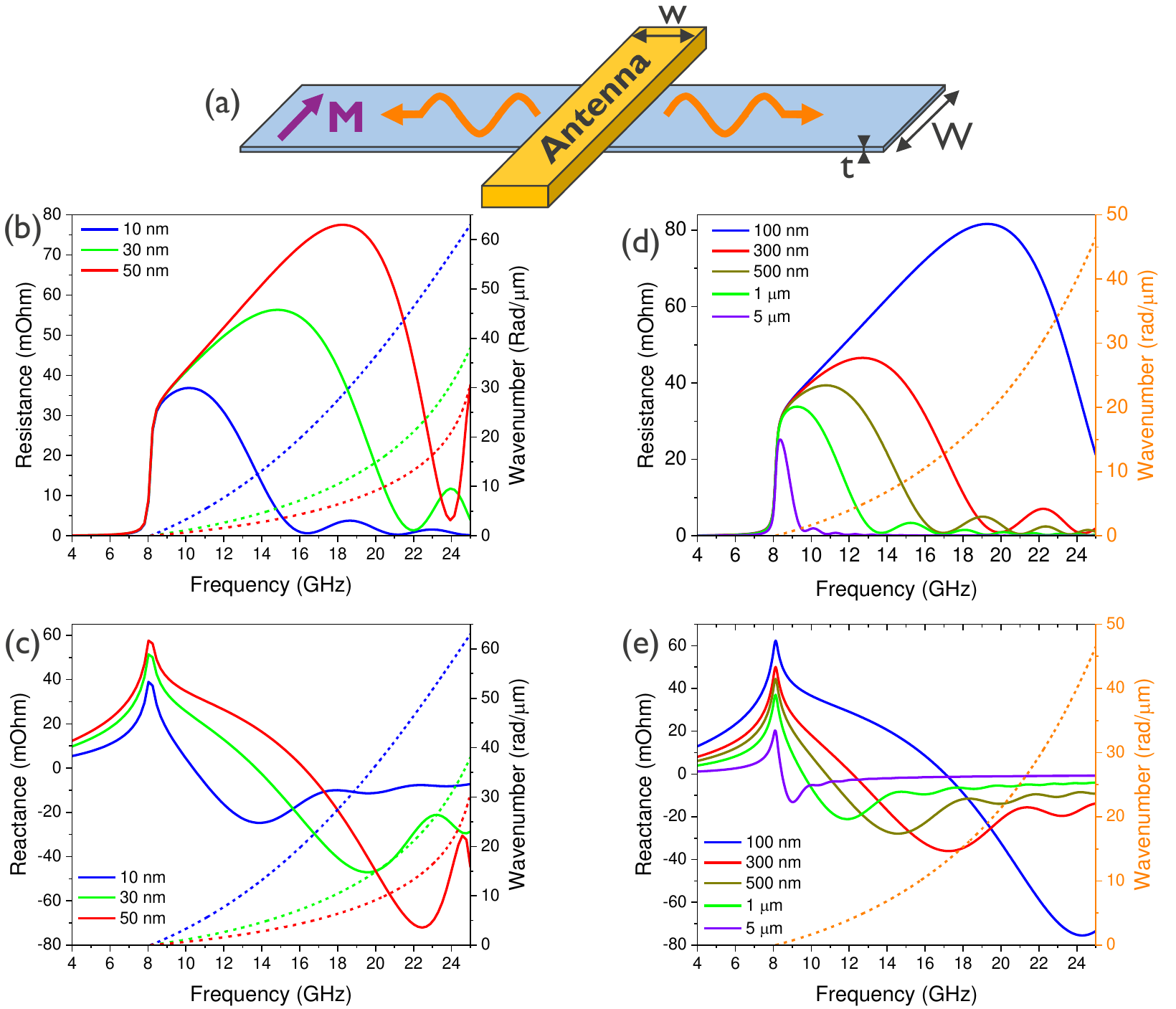}
	\caption{Dependence of the spin-wave impedance on the frequency $f$ in the \textbf{(a)} Damon--Eshbach configuration for different geometrical antenna and waveguide parameters. \textbf{(b)} Dependence of the  spin-wave resistance and \textbf{(c)} spin-wave reactance on the magnetic waveguide thickness $t$ ($W = 1$ \textmu{}m, $w = 300$ nm). \textbf{(d)} Dependence of the spin-wave resistance and \textbf{(e)} spin-wave reactance on the antenna width $w$ ($W = 1$ \textmu{}m, $t = 30$ nm). The FMR frequency is located at zero wavenumber, \emph{i.e.} at a frequency of 8.1~GHz for the parameters considered.}
	\label{fig:scaling}
\end{figure*}

By contrast, the influence of the magnetic waveguide thickness $t$ on the spin-wave impedance is more complex. On one hand, Eq.~\eqref{eq:Lm} contains a prefactor $tW$, which indicates that the spin-wave impedance is (linearly) increasing with increasing magnetic volume (and therefore also with magnetic waveguide thickness $t$). However, the thickness $t$ also strongly influences the spin-wave dispersion relation and therefore the spin-wave susceptibility. Figure~\ref{fig:scaling}(b) depicts the spin-wave resistance in the Damon--Eshbach geometry (\emph{cf.} Fig. \ref{fig:scaling}(a)) for three different magnetic waveguide thicknesses ($W = 1$ \textmu{}m, $w = 300$ nm). The data clearly indicate that a larger film thickness results in a larger spin-wave resistance. As mentioned above, this can be partly ascribed to a larger magnetic volume, which can store more magnetic power when the film thickness is increased. In addition, $t$ also affects the spin-wave dispersion, leading to a larger spin-wave group velocity (a steeper slope of the dispersion relation) for thicker films. This means that for a given frequency, the corresponding spin-wave wavenumbers are lower for thicker waveguides. Since the driving Oersted field spectrum has larger values at lower wavenumbers, this results in an additional increase of the spin-wave resistance for thicker waveguides. Similar conclusions can be drawn for the spin-wave reactance in Fig. \ref{fig:scaling}(c). Here, higher waveguide thicknesses also result in increased spin-wave reactance values originating from the increased magnetic volume and the steeper dispersion relation for thicker waveguides. 

A third important parameter is the antenna width $w$. Whereas $w$ does not affect the spin-wave dispersion relation and susceptibility, it modifies the driving Oersted field spectrum that determines the overlap integral with the magnetization dynamics in Eq.~\eqref{eq:Lm}. Figure~\ref{fig:scaling}(d) shows the spin-wave resistance for five different antenna widths ($W = 1$ \textmu{}m, $t =$ 30 nm). The data indicate that a smaller antenna width results in a larger peak spin-wave resistance. This can be understood by considering that the sinc-function that describes the Oersted field spectrum becomes broader for smaller antenna widths. For a wide antenna, the sinc-function strongly decays already at small wavenumbers, resulting in a reduced width of the spin-wave resistance peak and, in general, in smaller resistances for higher frequencies. A similar reasoning can be made for the spin-wave reactance in Fig.~\ref{fig:scaling}(e). Here, larger antenna width also results in smaller peaks due to the narrower antenna spectrum.

Besides geometrical parameters, the spin-wave impedance also depends on the properties of the waveguide material. The saturation magnetization $M_s$ is important as it strongly influences the FMR frequency and the spin-wave dispersion relation. A higher saturation magnetization leads to a higher the FMR frequency, thereby shifting the impedance curve to higher frequencies and higher values due to the proportionality to $\omega$. In addition, an increased $M_s$ also results in a steeper dispersion relation, analogously to an increased waveguide thickness, as discussed above. Thus, the effect of an increased $M_s$ is also similar to the effect of increasing the waveguide thickness, resulting in higher and broader impedance peaks for higher saturation magnetization values. Another important material parameter is the magnetic damping $\alpha$, which influences the magnetic susceptibility. A larger $\alpha$ results in a broader susceptibility spectrum and a smaller susceptibility peak value. Despite the reduction of the susceptibility peak value, this does not automatically result in a reduced impedance since the magnetic inductance depends on the spectral overlap integral between the susceptibility and the Oersted field (\emph{cf.} Eq.~\eqref{eq:Lm}). Hence, the broader the susceptibility and the Oersted field spectrum, the higher the overlap integral and thus the spin-wave impedance. 

\section{Power transmission efficiency of an inductive wire antenna transducer on a ferromagnetic narrow waveguide}

Above, we have discussed the frequency dependence of the spin-wave impedance on waveguide and antenna geometry, as well as on the magnetic material properties. From a spintronic application point of view, a key parameter is the power coupling efficiency that describes the ratio of the power that is transferred to the magnetic system and the total microwave power that is incident on the antenna transducer.

In a microwave circuit, the power transmission from a source into a load (here the inductive antenna) is determined by the matching conditions between source and load impedances. Different matching approaches exist. Maximum power dissipation in the load is obtained when the antenna impedance is equal to the complex conjugate of the source impedance $Z_\mathrm{eq} = Z^*_S$.\cite{Lee04} Note that, in this case, the maximum transferred power is half of the source power. Zero reflection at the load occurs for  $Z_\mathrm{eq} = Z_S$. By contrast, the maximum power transmission \emph{efficiency} is obtained for effective open circuit conditions, \emph{i.e.} for $R_S \ll R_\mathrm{eq}$, reaching an efficiency of 1 when $R_S/R_\mathrm{eq} \rightarrow 0$.

In practice, experiments often utilize a source with a real 50 $\Omega$ or 75 $\Omega$ source resistance, although this no necessary condition. A detailed discussion on impedance matching to common 50 $\Omega$ or 75 $\Omega$ microwave instrumentation is beyond the scope of the article (see \emph{e.g.} Refs. \onlinecite{Dong20, Razavi11, Bowick07}) and has been recently addressed for different inductive antenna designs.\cite{Connelly2021} However, some general power coupling considerations can be made based on the equivalent circuit in Fig.~\ref{fig:eq}, which allow for the discussion of the scaling properties of inductive antennas and the associated power coupling scaling. The equivalent circuit indicates that power can be dissipated both by emission of spin-waves or excitation of FMR ($P_m$) as well as by Ohmic losses in the antenna ($P_\Omega$). Both losses are proportional to the square of the microwave current in the antenna and thus their ratio does not depend on the total power dissipated in the antenna and thus also not on the detailed impedance matching approach. Therefore, the power transfer efficiency into the spin-wave system can be expressed as 
\begin{equation}
\eta = \zeta\frac{P_m}{P_\Omega + P_m} = \zeta\frac{R_m I^2_{rf}}{R_\Omega I^2_{rf} + R_m I^2_{rf}} = \zeta\frac{R_m}{R_\Omega + R_m}\,.
\label{eta}
\end{equation}
Here, $0 \le \zeta \le 1$ represents the power coupling efficiency into the antenna with respect to the source power, which depends on the matching conditions. Under complex conjugate matching conditions (maximum power transfer) or for reflectionless matching to a real source impedance, $\zeta = \frac{1}{2}$, whereas $\zeta = 1$ near effective open circuit conditions. However, in any case, the antenna transduction efficiency is limited by the ratio of the spin-wave and Ohmic resistances, given by Eqs.~\eqref{eq:Rm} and \eqref{eq:Rs}, respectively. The scaling behavior of the power transmission efficiency of an inductive spin-wave antenna depends thus on the scaling of its Ohmic resistance as well as of its spin-wave impedance, as discussed in the previous section.

\begin{figure}[tb]
	\includegraphics[width=16cm]{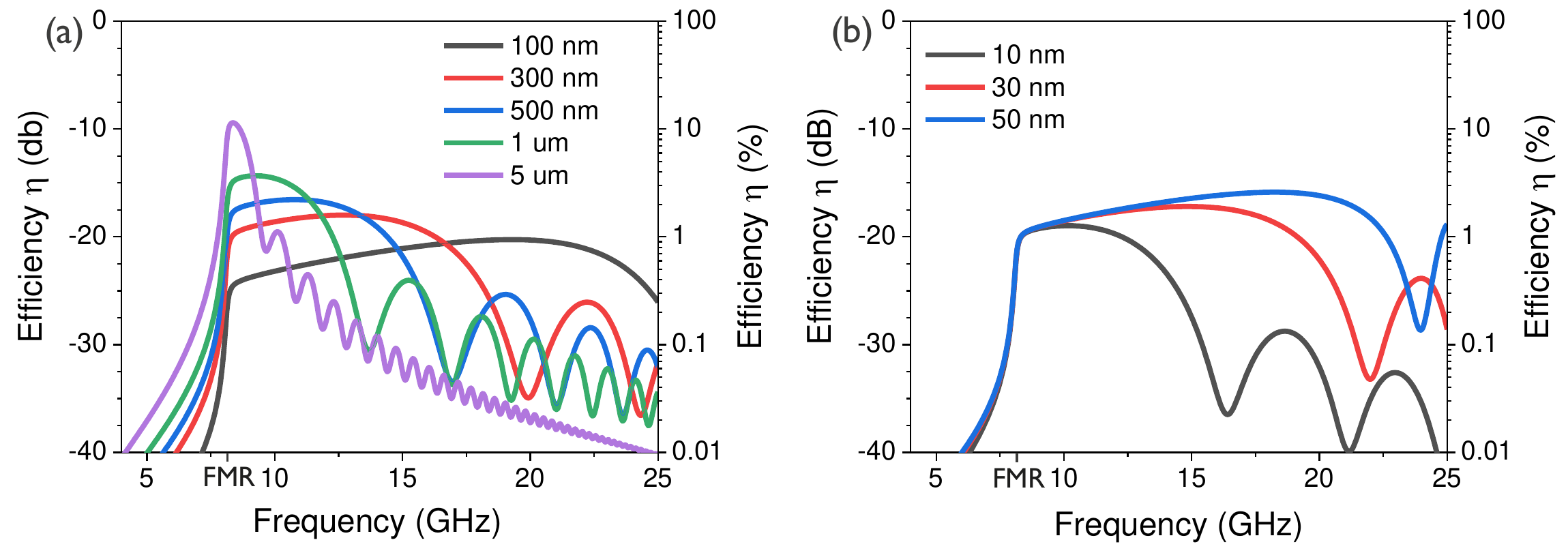}
	\caption{Power transmission efficiency $\eta$ into the magnetic system ($\zeta = \frac{1}{2}$, maximum power transfer) in the Damon--Eshbach configuration (a) as a function of the inductive antenna width $w$ between 100 nm and 5 \textmu{}m ($W = 1$ \textmu{}m, $t = 20$ nm) and (b) as a function of waveguide thickness $t$ between 10 nm and 50 nm ($W = 1$ \textmu{}m, $w = 300$ nm). The FMR frequency at 8.1 GHz is indicated in both graphs as a reference.}
	\label{fig:Gamma}
\end{figure}

We now quantitatively assess $\eta$ in the Damon--Eshbach configuration for an antenna thickness of $d$ = 40 nm, using the Cu resistivity of $\rho$ = 17 n$\Omega$m, as well as the above CoFeB material parameters. We further assume that the wire antenna length $\ell$ is equal to the waveguide width $W$, which is a lower limit for experimental realizations, as well as $\zeta = \frac{1}{2}$ (maximum power transfer or reflectionless matching to real source power). Figure~\ref{fig:Gamma}(a) shows $\eta$ for different antenna widths $w$. The data indicate that the maximum power coupling efficiency follows the spectral behavior of the spin-wave radiation resistance $R_m$. Note that the Ohmic antenna resistance $R_\Omega$ is assumed to be frequency independent as the skin effect is neglected. Although Fig.~\ref{fig:scaling}(b) shows that the spin-wave radiation resistance $R_m$ increases with decreasing $w$, especially at high frequencies, the Ohmic antenna resistance $R_\Omega$ also increases with decreasing $w$. As a consequence, the power transfer efficiency $\eta$ decreases with decreasing $w$ due to a faster increase of $R_\Omega$, despite the larger bandwidth. By contrast, the magnetic waveguide thickness $t$ only affects $R_m$ and therefore a larger $t$ not only increases $R_m$ but also improves $\eta$. Similar effects can be found for the dependence on the magnetic material properties, which can be optimized to enhance $R_m$ and therefore $\eta$. By contrast, changing the waveguide width or the antenna length does not alter the efficiency $\eta$ because the spin-wave and Ohmic resistance are equally proportional to both dimensions. For the chosen parameters and nanoscale dimensions, the maximum power transfer efficiency into the spin-wave system is of the order of 1 to 3\%{} ($-20$ to $-15$ dB).

These results quantitatively illustrate the scaling behavior of inductive antennas as spin-wave transducers. As observed in many experiments, the power transfer efficiency typically decreases when the device dimensions are reduced, in particular the antenna width as well as the waveguide thickness. In practice, poor matching between the source impedance and the equivalent antenna impedance $Z_\mathrm{eq}$ in Eq.~\eqref{eq:z_antenna} further reduce the power emitted into the spin-wave system. The results in the previous section show that absolute spin-wave radiation resistances $R_m$ are of the order of m$\Omega$. Nonetheless, for the given CoFeB materials parameters, the results indicate that high power transfer efficiencies $\eta$ can be obtained for \textmu{}m dimensions (antenna widths) when source and antenna impedances can be matched, \emph{i.e.} for $Z_\mathrm{eq} = Z^*_S$, in agreement with a recent report.\cite{Connelly2021} For example, $\eta$ can be as high as 11\%{} ($-9.5$ dB) for $W = \ell = 5$ \textmu{}m, $w =300$ \textmu{}m, and $t = 20$ nm. However, when the antenna width is scaled to nm dimensions, $\eta$ is reduced to 1--3\%{} ($-20$ to $-15$ dB) and below, indicating that wire antennas become increasingly inefficient when the bandwidth is increased by reducing their width. 

\section{Conclusions}

In conclusion, we have derived an equivalent circuit for inductive antennas as transducers between microwave currents and FMR or spin waves. Such antennas have been used commonly in magnonic experiments. Furthermore, we have derived analytical equations, assuming an arbitrary antenna shape as well as an arbitrary current density, for the different components of the equivalent circuit, which comprises an Ohmic resistance $R_\Omega$, a self-inductance $L_0$, as well as an additional complex partial inductance $L_m$ that stems from the coupling to the magnetic system. In addition, both exchange and dipolar interaction have been considered to describe the magnetization dynamics. The model has then been used to present a case study for a thin CoFeB waveguide with a straight wire antenna on top. Both the spin-wave resistance and reactance have been determined and general spectral and geometrical trends have been identified. The results have been used to assess the maximum transduction power efficiency for such a system, with a focus on its scaling behavior.

These results have multiple implications when magnonic devices including inductive antennas are scaled to nm dimensions. In real-world applications, the power transfer efficiency is a key parameter to determine the performance of any magnonic device. As shown above, spin-wave radiation resistances for sub-\textmu{}m straight wire antenna dimensions are of the order of a few 10 m$\Omega$. When the antenna width (and thickness) is scaled, the Ohmic resistance can become rapidly an order of magnitude or more larger, leading to a reduced power transfer efficiency and to more relative Ohmic power dissipation. In addition, the matching of small antenna impedances to conventional 50 $\Omega$ or 75 $\Omega$ sources can be challenging and reduces the available bandwidth considerably. The matching can be improved by increasing, \emph{e.g.} the waveguide width and the antenna length.\citep{Connelly2021} While this does not affect the power transfer efficiency, it increases the overall resistance (both Ohmic and spin-wave) and can be used to match the load impedance better to \emph{e.g.} 50 $\Omega$ source impedance; however, this also increases the structure (device) size. We note that analogous arguments apply to the case of the two-antenna mutual inductance introduced in Sec.~\ref{mutual} and the associated equivalent two-port networks.

As mentioned before, the spin-wave impedance is strongly determined by the spectral overlap integral between the magnetic susceptibility and the magnetic Oersted field. In the above discussion, a rectangular wire has been considered which results in a sinc-like Oersted field in reciprocal space. It is also possible to design different antenna shapes such as U-shaped antennas or IDTs, \emph{i.e.} multiple wire antennas, or envisage even more exotic designs, which improve the spin-wave inductance $L_m$ with respect to the straight wire case considered here.\citep{Connelly2021} In all such cases, the above derived model remains valid and can be applied to study the maximum efficiency of a particular antenna design and its scalability. The only parameter that needs to be modified is the spectrum of the Oersted in the overlap integral. The equivalent circuit and the model can thus be used to determine the coupling efficiency between the microwave source and the antenna (by impedance matching) as well as between the antenna and the magnetization dynamics (FMR, spin waves). While it is beyond the scope of the paper to discuss the particular spin-wave impedance value for the different antenna designs, the general trends and thinking strategies outlined above remain valid for different types of antenna transducers. Therefore, this model can be used as the starting point for the design and optimization of antenna transducers for (nano-)magnonic devices.

\begin{widetext}
\appendix
\section{Dynamic susceptibility}
\label{app:A}

The magnetization dynamics can be described by the LLG equation \cite{LANDAU92,Gilbert04}
\begin{equation}
	\label{eq:llg}
	\frac{d\mathbf{M}}{dt} = -\gamma_0(\mathbf{M} \times \mathbf{H}_{\mathrm{eff}}) + \frac{\alpha}{M_\mathrm{a}} \left( \mathbf{M} \times \frac{d\mathbf{M}}{dt} \right) \, ,
\end{equation}
\noindent with $\bm{M}=\bm{M}_0+\bm{m}$, $\gamma_0=|\gamma|\mu_0$, $|\gamma|$ the absolute value of the gyromagnetic ratio, $\bm{H}_\mathrm{eff}$ the effective magnetic field and $\alpha$ the magnetic damping constant. $\bm{M}_0$ and $\bm{m}$ are the static and dynamic components of the magnetization $\mathbf{M}$, respectively. In this work, the effective field consist of a static bias field $\bm{H}_0$, a dynamic antenna field $\bm{h}_\mathrm{a}$, spin-wave dipolar field $\bm{h}_\mathrm{d}$ and spin-wave exchange field $\bm{h}_\mathrm{ex}$. For weak magnetization dynamics, \emph{i.e.} $|\bm{m}|\ll |\bm{M}_0|$, the LLG equation can be linearized and becomes 
\begin{align}
	\nonumber	i\omega \bm{m}(\bm{k},\omega) = &-\gamma_0 \left( \bm{M}_0\times \left[\bm{h}_\mathrm{d}(\bm{k},\omega)+\bm{h}_\mathrm{ex}(\bm{k},\omega)\right] + \bm{M}_0\times \bm{h}_\mathrm{a}(\bm{k},\omega) +  \bm{m}(\bm{k},\omega)\times\bm{H}_0  \right) \\ 
		&   + \frac{i\omega \alpha}{M_0}\left( \bm{M}_0 \times \bm{m}(\bm{k},\omega)\right) \, ,
\end{align}
\noindent where complex notation was used and $\bm{k}$ and $\omega$ denote the spin-wave wavevector and angular frequency, respectively. The dynamic dipolar magnetic field is given by \cite{Gurevich96}
\begin{equation}
\bm{h}_\mathrm{d}(\bm{r}) = \int\limits_V \hat{\Gamma}(\bm{r},\bm{r}') \bm{m}(\bm{r},\bm{r}')\, d\bm{r}' \, ,
\end{equation}
\noindent with $V$ the volume of the magnetic material and $\hat{\Gamma}(\bm{r},\bm{r}')$ the magnetostatic Green's function given by 
\begin{equation}
\hat{\Gamma}(\bm{r},\bm{r}') = - \nabla_\mathbf{r} \nabla_\mathbf{r'} \frac{1}{|\mathbf{r}-\mathbf{r}'|}\,.
\end{equation} 
For a plane wave in a thin film, the averaged dipolar magnetic field inside the film becomes \cite{Damon60,Harte68,Gurevich96}
\begin{equation}
\mathbf{h}_\mathrm{d}(\mathbf{k},\omega) = - \begin{bmatrix}
P\sin^2(\theta) & P\cos(\theta)\sin(\theta) & 0  \\
P\cos(\theta)\sin(\theta) & P\cos^2(\theta) & 0 \\ 
0 & 0 &  1-P
\end{bmatrix} \mathbf{m}(\mathbf{k},\omega)\, ,
\end{equation}
\noindent with $\theta$ the (in-plane) angle between the static magnetization and the wavevector and
\begin{equation}
\label{eq:P}
P = 1-\frac{1-e^{-kt}}{kt} \, .
\end{equation}
The dynamic exchange field is given by
\begin{equation}
\mathbf{h}_{\mathrm{ex}}(\mathbf{k},\omega) = \lambda_{\mathrm{ex}} \nabla^2 \bm{m}(\bm{r}) \,.
\end{equation}
\noindent For a plane wave, this becomes
\begin{equation}
\mathbf{h}_{\mathrm{ex}}(\mathbf{k},\omega) = -\lambda_{\mathrm{ex}} k^2 \mathbf{m}(\mathbf{k},\omega)\, , 
\end{equation}
\noindent with 
\begin{equation}
	\lambda_\mathrm{ex} = \sqrt{\frac{2A_\mathrm{ex}}{\mu_0M_\mathrm{a}^2}}
\end{equation}
\noindent and $A_\mathrm{ex}$ the exchange stiffness constant. To simplify the expressions, the following tensor is introduced
\begin{equation}
	\hat{F} = \begin{bmatrix}
	P\sin^2(\theta)+\lambda_{\mathrm{ex}} k^2 & P\cos(\theta)\sin(\theta) & 0  \\
	P\cos(\theta)\sin(\theta) & P\cos^2(\theta)+\lambda_{\mathrm{ex}} k^2 & 0 \\ 
	0 & 0 &  1-P+\lambda_{\mathrm{ex}} k^2
	\end{bmatrix}\,.
\end{equation}
The general form of the linearized LLG equation then becomes 
\begin{equation}
	\bm{m} = \hat{\chi}_\omega  \bm{h}_{a} 
\end{equation}
with 
\begin{align}
	\hat{\chi}_\omega & = \\
\nonumber	& \Scale[0.81]{-\omega_M \begin{bmatrix}
	i\omega+\omega_MF_{xy} \zeta_z & \left[ \omega_0+\omega_MF_{yy}+i\omega\alpha \right]\zeta_z & -\left[ \omega_0+\omega_MF_{zz}+i\omega\alpha \right]\zeta_y \\
	\left[ \omega_0+\omega_MF_{xx}+i\omega\alpha \right]\zeta_z &  i\omega+\omega_MF_{xy} \zeta_z & \left[ \omega_0+\omega_MF_{zz}+i\omega\alpha \right]\zeta_x \\
	\left[ \omega_0+\omega_MF_{xx}+i\omega\alpha \right]\zeta_y - \omega_MF_{xy}\zeta_x & -\left[ \omega_0+\omega_MF_{yy}+i\omega\alpha \right]\zeta_x - \omega_MF_{xy}\zeta_y & i\omega
	\end{bmatrix}^{-1}}\,,
\end{align}
$\omega_M=\gamma_0M_0$, $\omega_0=\gamma_0H_0$, and
\begin{equation}
	\zeta = \begin{bmatrix}
	\cos(\psi) \cos(\theta) \\
	\cos(\psi)\cos(\theta) \\
	\sin(\psi)
	\end{bmatrix}\, .
\end{equation}
\noindent Here, $\psi$ is the angle between the static magnetization $\bm{M}_0$ and the film. When the static magnetization direction coincides with one of the coordinate axes, the expression can be simplified and written as
\begin{equation}
	\begin{bmatrix}
	\omega_1 +i\omega\alpha & -i\omega \\
	i\omega & \omega_2 +i\omega\alpha 
	\end{bmatrix} \begin{bmatrix}
	m_k \\ m_l
	\end{bmatrix} = \omega_\mathrm{M} \begin{bmatrix}
	h_k \\ h_l
	\end{bmatrix}\, ,
\end{equation}
which becomes
\begin{align}
	\begin{bmatrix}
	m_k \\ m_l
	\end{bmatrix} &= \hat{\chi}_\omega  \begin{bmatrix}
	h_k \\ h_l
	\end{bmatrix} \\
	&= \frac{\omega_M}{\left( \omega_1\omega_2-\omega^2\right)+i\alpha\omega\left( \omega_1+\omega_2 \right)} \begin{bmatrix}
	\omega_2+i\alpha\omega & i\omega \\
	-i\omega & \omega_1+i\alpha\omega
	\end{bmatrix} \begin{bmatrix}
	h_k \\ h_l
	\end{bmatrix}\\
	&= \frac{\hat{\Omega}^2}{\omega_\mathrm{r}^2-\omega^2+i\omega \Gamma}\begin{bmatrix}
	h_k \\ h_l
	\end{bmatrix}\, ,
\end{align}
\noindent with $\omega_\mathrm{r}=\sqrt{\omega_1\omega_2}$ the spin-wave resonance frequency or dispersion relation and $\Gamma=\alpha\left( \omega_1+\omega_2\right)$ the spin-wave damping rate. The factors $\omega_1$ and $\omega_2$ are configuration dependent and are given for the three specific cases below. The susceptibility can be separated into a real and imaginary part as given below
\begin{equation}
\hat{\chi}_{kk} = \frac{\omega_M(\omega_2(\omega_r^2-\omega^2) + \alpha^2\omega^2(\omega_1+\omega_2))}{(\omega_\mathrm{r}^2-\omega^2)^2+\omega^2 \alpha^2 (\omega_1+\omega_2)^2} - i \frac{\alpha\omega\omega_M(\omega^2+\omega_2^2)}{(\omega_\mathrm{r}^2-\omega^2)^2+\omega^2 \alpha^2 (\omega_1+\omega_2)^2}
\end{equation}
\begin{equation}
\hat{\chi}_{ll} = \frac{\omega_M(\omega_1(\omega_r^2-\omega^2) + \alpha^2\omega^2(\omega_1+\omega_2))}{(\omega_\mathrm{r}^2-\omega^2)^2+\omega^2 \alpha^2 (\omega_1+\omega_2)^2} - i \frac{\alpha\omega\omega_M(\omega^2+\omega_1^2)}{(\omega_\mathrm{r}^2-\omega^2)^2+\omega^2 \alpha^2 (\omega_1+\omega_2)^2}\,.
\end{equation}
The group velocity is defined as
\begin{equation}
	v_g = \frac{\partial \omega_r}{\partial k} = \frac{\partial \sqrt{\omega_1\omega_2}}{\partial k} = \frac{1}{2\omega_1\omega_2}\left[ \omega_2\frac{\partial \omega_1}{\partial k} + \omega_1\frac{\partial \omega_2}{\partial k}\right]\,.
\end{equation}

\subsection{Specfic cases}

For $\theta=0$ and $\psi=0$ (backward-volume configuration), $k=y$, $l=z$ and
\[
\begin{cases}
\omega_1 = \omega_0 +\omega_M \left( \lambda_\mathrm{ex}k^2 \right)\\
\omega_2 = \omega_0 +\omega_M \left( \lambda_\mathrm{ex}k^2+1-P\right)
\end{cases}\,.
\]

For $\theta=\pi/2$ and $\psi=0$ (Damon--Eshbach configuration), $k=z$, $l=x$ and
\[
\begin{cases}
\omega_1 = \omega_0 +\omega_M \left( \lambda_\mathrm{ex}k^2 + 1-P\right)\\
\omega_2 = \omega_0 +\omega_M \left( \lambda_\mathrm{ex}k^2+P  \right)
\end{cases}\,.
\]

For $\phi=\pi/2$ (forward-volume configuration), $k=x$, $l=y$ and
\[
\begin{cases}
\omega_1 = \omega_0 +\omega_M \left( \lambda_\mathrm{ex}k^2 +P \right)\\
\omega_2 = \omega_0 +\omega_M \left( \lambda_\mathrm{ex}k^2 \right)
\end{cases}\,.
\]

\end{widetext}

\begin{acknowledgments}
This work has received funding from the European Union’s Horizon 2020 research and innovation program within the European Innovation Council Pathfinder FET-OPEN project CHIRON under grant agreement No.~801055. It has also been supported by imec’s industrial affiliate program on beyond-CMOS logic. F.V. acknowledges support by the Research Foundation -- Flanders (FWO) through a PhD fellowship.
\end{acknowledgments}

\section{Author contributions}
F.V. and V.T. developed the theoretical model. F.V., F.C. and C.A. described the scaling behavior. G.T. and B.S. gave additional insight and critical notes. All authors reviewed the manuscript and contributed to the main conclusions.

\clearpage

\end{document}